\documentstyle[12pt]{article}
\input{psfig}
\newcommand{\bm}[1]{\mbox{\boldmath $#1$}}
\def\be{\begin{equation}}                
\def\ee{\end{equation}}
\def\Vb{{\bm V}_b}
\def\vel{{\bm v}}

\def\Thzero{\Theta^{(0)}}
\def\Thuno{\Theta^{(1)}}
\def\chiv{{\bm \chi}}
\def\Xiv{{\bm \Xi}}
\def\Upsv{{\bm \Upsilon}}

\title{Scalar transport in compressible flow}
\author{M. Vergassola$^{1}$ and M. Avellaneda$^{2}$}
\begin{document}
\maketitle
\centerline{$^1$ CNRS, URA 1362, Observatoire de Nice, B.P. 4229,
06304 Nice Cedex 4, France.}
\centerline{$^{2}$ Courant Institute of Mathematical Sciences, New York
University, New York, N.Y. 10012.}
\date{}
\medskip
\vskip1.5cm
\begin{abstract}
Transport of scalar fields in compressible flow is investigated.
The effective equations governing the transport at scales large compared
to those of the advecting flow $\vel$ are derived by using multi-scale 
techniques. Ballistic transport generally takes place 
when both the solenoidal and the potential components
of $\vel$ do not vanish, 
despite of the fact that $\vel$ has zero average value. The calculation
of the effective ballistic velocity $\Vb$ is reduced to the solution of one
auxiliary equation. An analytic expression for $\Vb$ is derived in some 
special instances, i.e. flows depending on a single coordinate,
random with short correlation times and slightly
compressible cellular flow. The effective mean velocity $\Vb$ vanishes 
for velocity fields which are either incompressible or potential and 
time-independent. For generic compressible flow, the most general conditions
ensuring the absence of ballistic transport are isotropy and/or
parity invariance. When $\Vb$ vanishes (or in the frame of reference
comoving with velocity $\Vb$), standard diffusive transport takes place.
It is known that diffusion is always enhanced by incompressible flow.
On the contrary, we show that diffusion is depleted in the presence
of time-independent potential flow. Trapping effects due to
potential wells are responsible for this depletion. For time-dependent 
potential flow or generic compressible flow, transport rates are
enhanced or depleted depending on the detailed structure of the 
velocity field. 
\end{abstract}
\newpage

\section{Introduction}

One of the most interesting issues in statistical mechanics
is related to the effects of a given microscopic dynamics on 
phenomena occurring at much larger scales. A well-known example
is provided by kinetic theory. When the Knudsen number is small, i.e.
the scales of interest are much larger than the mean free path, hydrodynamic
equations can be derived. In these equations only the macroscopic scales
are involved. The microscopic small-scale degrees of freedom are averaged
out and do not appear explicitly. This important point is physically
related to the fact that macroscopic and microscopic space-time scales
are widely separated (scale separation). In this situation it is indeed
just the mean cumulative result of many almost independent effects which
is relevant for the large-scale dynamics. This also explains the universality
of hydrodynamic equations, whose structure is essentially dictated by the
symmetries and the conservation laws of the microscopic dynamics. The details
of the dynamics only affect the numerical value of the transport coefficients,
e.g. the viscosity in Navier-Stokes equations.
A remarkable application of this universality has been made in 
lattice gas methods \cite{FHP,Complex}.

The analogy with kinetic theory is very useful 
for any system involving scale separation. Let us consider 
for example the transport of a scalar field $\Theta$ by a 
velocity field $\vel$ \cite{HKM}. 
Such a problem arises, for example,
studying the spreading of particles 
advected by a given velocity field and subject to
molecular diffusion. 
At space-time scales much larger 
than those of $\vel$, we physically expect the dynamics of $\Theta$ 
to be governed by
an effective transport equation. From a technical point of view, 
exploiting the scale separation
to derive the effective equations involves a singular perturbation 
problem (see, e.g., Ref.~\cite{BO}). To deal with it, multi-scale techniques
(also known as homogenization) have been mostly used \cite{BLP}.
A well known application of these techniques is for scalar transport 
by incompressible
velocity fields (divergence-free). In Ref.~\cite{Piretal}
it was shown that the large-scale dynamics of the scalar in the
presence of scale separation is indeed governed by an effective equation,
which is always diffusive. 
The calculation of the effective diffusivity is reduced 
to the solution of one auxiliary equation. From this equation 
it follows that
the effective diffusivity is always larger
than the molecular one, i.e. incompressible flow enhance diffusion.
The auxiliary equation can be solved and the effective diffusivity 
calculated exactly in 
some special instances, e.g. parallel flow \cite{YZ}, or asymptotically in the 
limit of high P\'eclet numbers for cellular flow \cite{Shraiman,Rosetal}. 
Numerical techniques are otherwise generally needed \cite{BCVV,Majda}.
Variational principles \cite{FP} for the effective diffusivity
have also been derived both for static and time-dependent 
incompressible flow.
Note that for random flow arbitrarily small wavevectors might be excited.
The existence of a range of scales where
scale separation applies is thus not guaranteed. 
A necessary condition is that the variance of the vector potential
be finite \cite{AM,AV}. When the previous condition is violated,
anomalous transport might occur. Hereafter and in the 
sequel we shall suppose that, for scales sufficiently large, scale separation 
holds and we shall be interested in the dynamics at these scales.
Another transport phenomenon which has attracted much interest is momentum
transport in Navier-Stokes incompressible flow. 
This system provides a very interesting counterexample to the possible
belief that, if scale separation holds, the dynamics of the large scales
is necessarily diffusive. The so-called anisotropic
kinetic alpha effect \cite{FSS} generically appears in the absence of 
parity-invariance and isotropy. Furthermore, even when these two 
conditions are satisfied,
the effective viscosity might be negative \cite{Siv,GVF}. Large scales 
are then strongly amplified,   
nonlinearities become relevant and the dynamics at
large scales presents some peculiar properties \cite{MV,BMV}. 

The aim of our paper is to investigate scalar transport
in compressible flow. The results presented here illustrate the 
important differences between scalar transport by compressible
and incompressible flow. First, it is known that transport rates 
are always enhanced by incompressible flow. We show that for 
generic compressible flow this property is not valid. For 
arbitrary time-independent potential flow we 
actually prove that transport is always depleted because of trapping. 
For time-dependent potential flow or generic compressible flow 
it is shown by explicit examples that transport rates are
enhanced or depleted depending on the specific structure of the 
velocity field. Second, and more surprisingly, trapping effects
due to the compressibility of the flow can enhance 
the spreading of particles and 
lead to very efficient ballistic transport. This means that the average 
distance of a particle from its initial position does not vanish,
but actually grows linearly with time. Note that the small-scale 
velocity field is supposed to have zero average value. Still, because
of compressibility, an effective mean velocity emerges in the large-scale
dynamics. Particles are indeed strongly concentrated or rarefied 
and sample the small scales in a very non-uniform way. It is therefore 
possible to have an effective large-scale mean velocity 
even if the small-scale flow has zero average value.

The paper is organized as follows. In Section~\ref{s:formul}, we 
formulate the problem 
and present some simple heuristic arguments.
The multi-scale technique for ballistic transport is presented in
Section~\ref{s:ball}. Simple examples of flows leading to 
ballistic transport are discussed in Section~\ref{s:examples}. 
The general conditions ensuring the absence of the ballistic effect, i.e.
isotropy and/or parity invariance, are finally discussed in 
Section~\ref{s:absence}. For the flows where the ballistic 
effect is absent, transport is diffusive in the presence of scale separation.
The general formalism for the calculation of the effective diffusivities in 
compressible flow is presented in Section~\ref{s:diff}. The special case of
static potential flow is investigated in detail in Section~\ref{s:deplete}. 
The final Section is reserved for conclusions.

\section{Formulation and heuristics}
\label{s:formul}

Particles advected by a velocity field ${\bm v}$ and subject to molecular
diffusion obey the following Langevin equation
\be
\label{Langevin}
{d{\bm x}({\bm a},t)\over dt}\,=\, {\bm v}({\bm x},t) + {\bm W}(t).
\ee
The function ${\bm x}({\bm a},t)$ denotes the position at time $t$ of the 
particle which was initially in ${\bm a}$. 
The random process ${\bm W}(t)$ is Gaussian, 
independent of $\vel$, has zero mean and is white-noise in time 
\be
\label{whitenoise}
E\left(W_i(t)\,\, W_j(t')\right) = 2\,\kappa\,\delta_{ij}\, \delta(t-t').
\ee
The constant $\kappa$ appearing in (\ref{whitenoise}) is the molecular 
diffusivity. The velocity field ${\bm v}$
is here supposed to belong to one of the following classes\,:
(i) deterministic and periodic in space and time or periodic in space and 
time-independent. The period in the various 
directions need not be the same\,;
(ii) random, homogeneous and stationary or homogeneous and time-independent.
In both cases the velocity is a {\em prescribed} function of ${\bm x}$, and
possibly of $t$, and we shall not be concerned with 
the mechanisms maintaining
the flow. The mean value $\langle \cdot \rangle$ denotes the average 
over the periodicities for deterministic flow and the ensemble average 
in the random case. The velocity field has vanishing average value
\be
\label{mean}
\langle {\bm v}\rangle = 0.
\ee
Note that this hypothesis does not involve any restriction since we can 
always perform a Galilean transformation to reduce to the case (\ref{mean}).

The Fokker-Planck equation for the probability density $\Theta ({\bm x},t)$
which is associated with the Langevin equation
(\ref{Langevin}) is (see, e.g., Ref.~\cite{Risken})\,:
\be
\label{FP}
\partial_t\Theta + {\bm \partial}\cdot\left(\vel\Theta\right) = \kappa\,
\partial^2 \Theta,
\ee
where ${\bm \partial}$ denotes the spatial gradients. 
The problem discussed in
this paper is the large-scale behaviour of (\ref{FP}).
Specifically, we are interested in the dynamics of the field $\Theta$
on scales large compared with those typical of the small-scale dynamics, 
e.g., the periodicities of $\vel$. This is the situation encountered 
in the evolution of spots of particles {\em for large 
times}. As it was discussed in the introduction, the goal is to derive 
effective transport equations involving only the large-scale degrees of
freedom and to calculate the transport coefficients.

Before presenting the systematic formalism of the 
next Section, it is worth to consider 
a simple heuristic argument, in the same 
spirit as in Ref.~\cite{HKM78} for
the dynamo problem. Taking the average  
in (\ref{FP}) and trying to write down a
closed equation for the mean $\langle\Theta\rangle$, we face the classical 
closure problem\,: the Reynolds stress tensor $\langle
\vel\Theta\rangle$ cannot be
expressed as a function of $\langle\Theta\rangle$. The same closure problem
appears also analyzing higher-order fields correlations \cite{Israel}, 
which will not be considered here. However, 
in the presence of scale separation the situation changes.
An approach similar to the Born-Oppenheimer approximation in 
solid-state physics \cite{AschM} becomes feasible. 
There, the electrons are much faster and follow at any moment 
the ionic configuration. Here, the small-scale dynamics 
has enough time to readjust to the large-scale field and becomes 
essentially slaved to it.
It is then possible to express the Reynolds stress tensor by 
a gradient decomposition as
\be
\label{grad}
J_i = \langle v_i\Theta\rangle = (V_b)_i\langle\Theta\rangle 
- \kappa^E_{ij}\nabla_j
\langle \Theta\rangle + \ldots ,
\ee
where the remainder includes higher order terms in the gradients.
Inserting (\ref{grad}) into (\ref{FP}), we obtain
\be
\label{biblio}
\partial_t \langle\Theta\rangle + \Vb\cdot\nabla\langle\Theta\rangle=
\kappa^E_{ij}\nabla_j \nabla_j \langle \Theta\rangle .
\ee
This equation describes ballistic transport with mean velocity
$\Vb$ and diffusion in the comoving frame of reference with an
effective diffusivity $\kappa^E_{ij}$. A spot of particles 
will then be rigidly transported by the mean velocity $\Vb$ 
and at the same time
its contour is deformed, diffusing with diffusion coefficient 
$\kappa^E_{ij}$. The corresponding long-time behaviour 
of particles dispersion is
\be
\label{alla}
\langle (x_i(t)-x_i(0))\,(x_j(t)-x_j(0))\rangle = \left(V_b\right)_i\,
\left(V_b\right)_j\,t^2+2\kappa^E_{ij}\,t.
\ee
It is remarkable that, despite
of the zero average condition (\ref{mean}), a mean velocity $\Vb$ appears
in (\ref{biblio}) and the dispersion in (\ref{alla}) is quadratic in $t$.
The physical mechanisms leading to this effect
will be made clear by the examples of Section~\ref{s:examples}. There 
are nevertheless some classes of flow where we can already foresee 
(the systematic arguments are presented in Section~\ref{s:absence})
that the ballistic term in (\ref{biblio}) will be absent
and dispersion is linear in $t$. First of all,
for incompressible flows the term 
proportional to the mean $\langle\Theta\rangle$ should not appear.
A constant field,
being a trivial solution of (\ref{FP}), has in fact no dynamical role. 
Second, for parity-invariant or isotropic flows we also expect 
$\Vb=0$. A deterministic parity-invariant flow has indeed a center of
symmetry, e.g. the origin, so that
\be
\label{parity}
\vel(-{\bm x},t) = - \vel({\bm x},t).
\ee
For random flow, parity invariance means that the
statistical properties of $\vel$ are left invariant by the operations
${\bm x}\mapsto -{\bm x}$ and $\vel\mapsto -\vel$. 
Since any vector and its
opposite are equivalent, it follows $\Vb=0$. The effective velocity also 
vanishes for isotropic flow since no 
preferential direction can be picked out.

Note  that two other examples of transport phenomena also 
leading to a 
first-order dynamics are known\,: the anisotropic kinetic alpha 
(AKA) effect \cite{FSS} and the $\alpha$-effect \cite{HKM78} in
magneto-hydrodynamics. For the presence of both these effects 
the breaking of parity-invariance is needed
and the AKA effect disappears for isotropic flow as well. 
The main difference with respect to ballistic transport
is that the AKA and the $\alpha$ dynamics are not 
purely dispersive and lead to
large-scale instabilities in three dimensions.

\section{Multi-scale theory for ballistic transport}
\label{s:ball}

A general method for dealing with the singular perturbation problems 
encountered in systems with scale separation
is provided by multi-scale techniques \cite{BLP}.
As discussed in the previous Section, we are interested in the dynamics 
of the large scales. In concrete situations the scale separation between 
small and large scales is 
obviously always finite. The common procedure using asymptotic
methods is however to consider the limit of very small perturbations, obtain 
an expansion which is asymptotically valid and use it then for finite
values of the perturbation (see, e.g., Ref.~\cite{BO}). 
In our case the units are therefore 
defined in such a way that the typical scales
of the velocity field $\vel$ are $O(1)$, while
the large scales where transport takes place are supposed 
$O(\epsilon^{-1})$, with $\epsilon\ll 1$.
In multi-scale techniques a new set of space-time variables ${\bm X}$
and $T$ (called ``slow'') is introduced. The rationale in the choice 
of the variables is that the large-scale dynamics should take place on
scales $O(1)$ in the new variables. For example, since the spatial
large scales are $O(\epsilon^{-1})$, we should define   
${\bm X}=\epsilon {\bm x}$. As about the choice of the time variable,
it depends on the specific case. Ballistic transport corresponds to a
first-order equation in space and time. It follows that slow time for
this case should be defined as $T=\epsilon t$. On the contrary, for diffusive
transport the rescaling should be $T=\epsilon^2 t$ since the 
diffusion equation is first-order in time, but second-order in space. The
different re-scalings illustrate the fact, well-known in hydrodynamics,
that advection and diffusion take place on different time scales. In order
to capture both one could use a two-time formalism, as in \cite{Complex}.
However, for the sake of clarity we prefer treating advection and diffusion 
separately. In this Section, which is devoted to
ballistic transport, we define then $T=\epsilon t$.
The key for overcoming the singularity of the perturbation is to pretend
that fast and slow variables are independent. This allows to correctly
capture the dynamics of stirring between large and small-scale modes
which is missed by regular perturbation theory (see, e.g., Ref.~\cite{BO}).
It follows that
\be
\label{slow}
\partial_i \mapsto \partial_i + \epsilon \nabla_i \,\,\,,
\qquad \partial_t \mapsto \partial_t + \epsilon \partial_T \,\,\,,
\ee
where we shall denote the derivatives with respect to fast space variables
by the symbol $\partial$ and those with respect to slow variables
by $\nabla$. The solution $\Theta$ of (\ref{FP}) is then sought as a series
in $\epsilon$\,:
\be
\label{expansion}
\Theta = \Theta^{(0)}+\epsilon\, \Theta^{(1)} + \epsilon^2\, \Theta^{(2)} +
\ldots ,
\ee
where all functions depend {\it a priori} on both fast and 
slow variables. Let us then insert 
the expansion (\ref{expansion}) and the derivatives
(\ref{slow}) into the original equation (\ref{FP}). By equating terms having
equal powers of $\epsilon$, a hierarchy of equations is generated. 
The general structure of the equations is\,:
\be
\label{proto}
{\cal A} f \equiv 
\partial_t f + {\bm \partial}\cdot\left(\vel f \right) - \kappa\,
\partial^2 f = g,
\ee
where $f$ is the unknown function and $g$ is known (possibly as a function
of the solution of lower order equations). 
Since the operator ${\cal A}$ has derivatives
on the left of all the terms, 
the solvability conditions $\langle g\rangle =0$ must
be satisfied for equation (\ref{proto}) to have a solution 
(Fredholm alternative)\footnote{The solvability condition is, more
specifically, that $g$ should be orthogonal to the null space 
of the operator adjoint to ${\cal A}$, which is made of constants.
Note also that the same multi-scale methods could be used to analyze the
long-time behaviour of the
equation $\partial_t\theta + \vel\cdot{\bm \partial}\;\theta
= \kappa\,
\partial^2 \theta$. Its adjoint operator is closely related to the 
Fokker-Planck operator (\ref{FP}). This allows to carry over most 
multi-scale results valid for (\ref{FP}) to the previous equation.} 
The solvability conditions fix the dependence
of the $\Theta^{(n)}$ functions in (\ref{expansion}) on the slow variables
averaging over the small-scale degrees of freedom.
It is therefore not surprising that it is precisely by the 
solvability conditions that the effective equations  
for the dynamics of the large scales are obtained.

The equations which are relevant for the analysis of ballistic
transport are\,:
\begin{eqnarray}
\label{dueeqs1}
{\cal A} \Thzero &=& 0\, , \\
\label{dueeqs2}
{\cal A} \Thuno &=& -\partial_T \Thzero - \vel\cdot\nabla \Thzero
+2\kappa{\bm \partial}\cdot\nabla\Thzero .
\end{eqnarray}

Let us first briefly consider 
the incompressible case. Since ${\bm \partial}\cdot
\vel = 0$, the solution of (\ref{dueeqs1}) is a trivial constant field
$\Thzero = \Thzero({\bm X},T)$. Inserting this 
expression into (\ref{dueeqs2})
and using (\ref{mean}) and the periodicity of $\vel$ (or its homogeneity
in the random case), we easily find that 
\be
\label{incompressible}
\partial_T \langle\Thzero\rangle = 0 .
\ee
No dynamical process is therefore taking place on 
time-scales $O(\epsilon^{-1})$. This is in agreement with the known result 
\cite{Piretal} that
(in the presence of scale separation) 
incompressible flow always lead to diffusive transport, 
which involves time scales $O(\epsilon^{-2})$.

The situation changes when
a compressible velocity field is considered. Using the linearity
of (\ref{dueeqs1}), its solution $\Thzero$ can be expressed as
\be
\label{measure}
\Thzero = \langle \Thzero\rangle ({\bm X},T)\,\,m({\bm x},t) ,
\ee
where $m$ has unit average value. The solution $m$ is in general 
a nontrivial function due to the interplay between the solenoidal and the
potential components of $\vel$. By inserting 
(\ref{measure}) into (\ref{dueeqs2}), the following solvability condition 
is found
\be
\label{ballistic}
\partial_T \langle \Thzero\rangle ({\bm X},T) + \Vb\cdot\nabla \langle \Thzero
\rangle ({\bm X},T) = 0.
\ee
Equation (\ref{ballistic})
is indeed the one describing ballistic transport. The progress with 
respect to the heuristic arguments of  
Section~\ref{s:formul} is that we have now the expression of   
the effective velocity 
\be
\label{Vb}
\Vb = \langle m({\bm x},t)\,\vel ({\bm x},t)\rangle ,
\ee
with the field $m$ being the solution of the auxiliary problem
\be
\label{boh}
\partial_t\, m + {\bm \partial}\cdot\left(\vel\, m \right) - \kappa\,
\partial^2 \,m = 0\,;\qquad \langle m\rangle =1 .
\ee

Note that (\ref{Vb}) has a very 
simple physical interpretation. The interplay between potential 
and solenoidal components modifies an initially 
uniform distribution of particles
into a nontrivial density $m$. 
Equation (\ref{Vb}) simply illustrates the fact
that the small scales are sampled non-uniformly with the 
density $m$, which is naturally selected by the dynamics.

In Section~\ref{s:absence} we shall discuss the conditions ensuring
$\Vb=0$. For generic compressible flow $\Vb$ will however not vanish 
and we have reduced its calculation 
to the solution of (\ref{boh}). Except for 
a few cases, discussed in the sequel, 
this equation cannot be solved analytically and numerical
methods should be used. We shall not dwell here on numerical aspects
(see, e.g., Refs.~\cite{BCVV,Majda}). We just remark that solving 
auxiliary problems rather than the original equation is very
convenient since only small scales are involved in (\ref{boh}).

\section{Examples of ballistic transport}
\label{s:examples}

\subsection{Flows depending on a single coordinate}
\label{s:parallel}

The main features of ballistic transport are well illustrated by the
simple class of time-independent flow depending on a
single coordinate. For these flows we show here 
that the effective velocity
$\Vb$ can be calculated analytically. 

A two-dimensional example is 
\be
\label{stratified}
\vel({\bm x}) = \left(-{d\phi\over dx}(x)\,;\,w(x)\right).
\ee
The flow is clearly a superposition of the potential part $- d\phi / dx$
and the solenoidal component $w$. 
In the following we shall consider the two-dimensional case for
simplicity, but our results 
can be easily extended to more dimensions.
The solution of (\ref{boh}) for the flows (\ref{stratified})
is the Boltzmann distribution corresponding to the potential $\phi$\,:
\be
\label{Boltz}
m = {\exp\left(-\phi/\kappa\right)\over 
\langle\exp\left(-\phi/\kappa\right) \rangle }.
\ee
Note that (\ref{Boltz}) depends on the potential component only. 
This property is due to the very simple
geometry involved in the flows (\ref{stratified}).
It follows from (\ref{Boltz}) that 
the $x$-component of the ballistic velocity vanishes because of 
homogeneity or periodicity (see eq.~(\ref{merde})).
On the other hand, the expression of 
the $y$-component is\,:
\be
\label{crucco}
\left( V_b\right)_y = 
{\langle w(x)\exp\left(-\phi(x)/\kappa\right)\rangle
\over \langle \exp\left
(-\phi(x)/\kappa\right) \rangle },
\ee
which in general does not vanish. Equation (\ref{crucco}) provides the 
analytic expression of the effective ballistic velocity.
%-----------------------------------------------------------------
\setcounter{figure}{0}
\begin{figure}
\centerline{\psfig{figure=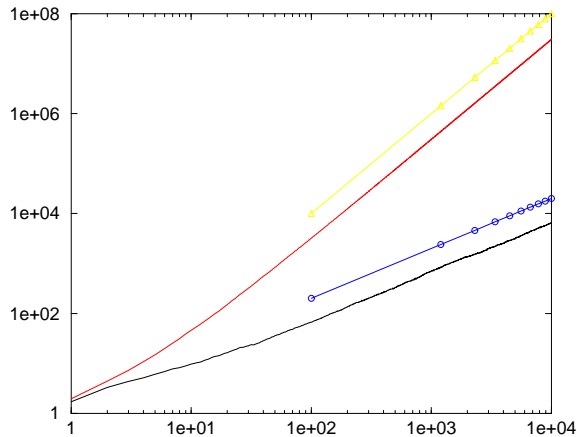,height=7cm}}
%,width=16cm}}
\vspace*{-0.7cm}
%\vspace{1cm}
\caption{Log-log plot of the dispersions
$\langle x^2(t)\rangle$ (lower curve) 
and $\langle y^2(t)\rangle$ corresponding to
the flow (22), versus time. 
These curves were obtained by Monte-Carlo simulation.
The asymptotic slopes are respectively 1 and 2, consistently with the
fact that the ballistic velocity points in the $y$-direction.
The values for the ballistic velocity generated in the simulation
agree with the theoretical prediction (23).}
\vspace{0.3cm}
\label{fig1}
\end{figure}
%-----------------------------------------------------------------

For the sake of concreteness, let us 
consider in more detail the following
simple periodic flow\,:
\be
\label{simple}
\vel = (\sin x\,;\,-\cos x).
\ee
The averages appearing in (\ref{crucco}) can be calculated in terms
of the Bessel functions $I_0$ and $I_1$ of imaginary argument. The
relevant formulae, from \cite{GR}, are listed for convenience in
Appendix~B. The final result is 
\be
\label{Bessel}
\left( V_b\right)_y = {I_1(1/\kappa) - I_1(-1/\kappa)\over 
I_0(1/\kappa) + I_0(-1/\kappa)}={I_1(1/\kappa)\over I_0(1/\kappa)}.
\ee
For large values of the molecular diffusivity $\kappa$, the Bessel 
functions can be expanded and
\be
\label{large}
\left( V_b\right)_y = {1\over 2\kappa} + O(1/\kappa^2),
\ee
which agrees with the general expression (\ref{primo}) 
derived in Appendix~A.
When $\kappa$ is small, we can use the asymptotic expansions (\ref{AppB5})
and (\ref{AppB6}) to obtain
\be
\label{small}
\left( V_b\right)_y = 1 - {\kappa\over 2} + O(\kappa^2).
\ee
Both (\ref{large}) and (\ref{small}) can obviously be also obtained 
directly from (\ref{crucco}). For small $\kappa$'s the 
field $m$ has a sharp maximum in $\pi$ around which it reduces
to a Gaussian with variance $\kappa$.
The result (\ref{small}) is then obtained by using Laplace method.
The expansion (\ref{small}) 
is in agreement with the fact that $w(\pi )=1$ and 
molecular noise forbids the concentration of all particles in the 
minimum of the potential. A closer look at the streamlines
of the flow reveals
the very simple dynamics 
associated with the ballistic transport (\ref{simple}).
The potential component 
concentrates the particles in the middle of
the channel $(0,2\pi)$. Here, the solenoidal 
part $w$ transports the particles in the positive $\hat{y}$-direction. 
In the general case the mechanism is more complicated and 
cannot be treated analytically
but this simple example highlights the essential role of 
both the solenoidal and the potential components for ballistic transport.

\subsection{A cellular flow}
\label{s:cellular}

Another velocity field such that the auxiliary equation 
(\ref{boh}) can also be tackled analytically, at least partially, 
is the following slightly perturbed cellular flow
\be
\label{perturbed}
\vel = (\cos y + \delta \cos x \,;\,\sin x)\equiv \vel^I + \hat{
\bm x} \delta
\cos x.
\ee
The term proportional to the constant $\delta$ is a 
small compressible perturbation, while
for $\delta=0$ the flow 
(\ref{perturbed}) reduces to the projection of the incompressible 
ABC flow \cite{Dombre} (with A=0, B=C=1) on the $x-y$ plane. We shall
derive here the asymptotic expression of the effective velocity $\Vb$ valid
for large P\'eclet numbers.

The incompressible BC flow $\vel^I$ consists of square convective cells
with separatrices parallel to the diagonals. The mechanism of effective 
diffusion 
at high P\'eclet numbers for this type of flow is well understood
\cite{Shraiman,Rosetal}. Particles within the cells simply circulate along
the streamlines and their density rapidly becomes almost uniform. 
On the contrary, 
particles close to the separatrices may cross the cell boundary
because of molecular noise. 
The width of the layers where these transitions 
typically occur is proportional to $\sqrt{\kappa}$. For small $\kappa$,
it is therefore possible to calculate the asymptotic behaviour of the 
effective diffusivity $\kappa^E$ by using boundary layer theory.
The stream function of our BC flow $\vel^I$ 
coincides with the one considered in Ref.~\cite{Rosetal} with
the transformations
\be
x'={x+y\over \sqrt{2}}\,;\qquad y'={x-y\over \sqrt{2}}-{\pi\over 2},
\ee
and $\tilde{u}=\sqrt{2}$, $d=\sqrt{2}\pi$,
$\beta=1$
(see eqs. (2) and (3) in Ref.~\cite{Rosetal}). The corresponding expression
of the effective diffusivity derived in Ref.~\cite{Rosetal} (eqs. 
(26) and (28)) is\,:
\be
\label{kappasqrt}
\kappa^E\simeq {2\over 0.749\sqrt{\pi}}\kappa^{1/2}.
\ee
%-----------------------------------------------------------------
\begin{figure}
\centerline{\psfig{figure=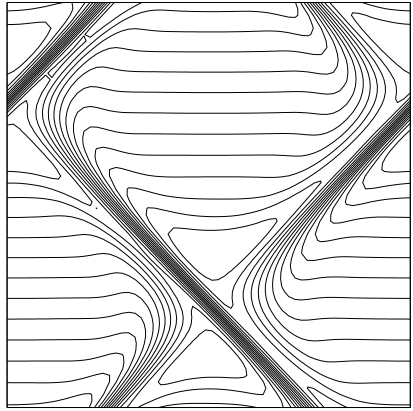,height=7cm}}
%,width=16cm}}
\vspace*{-0.7cm}
%\vspace{1cm}
\caption{Contour lines of the function $m$
corresponding to the 
slightly perturbed flow (26), obtained by solving the auxiliary
problem (18) numerically. 
The parameter values are $\delta=0.05$ 
and $\kappa=0.01$.}
\vspace{0.3cm}
\label{kappa0.01}
\end{figure}
%-----------------------------------------------------------------

On the other hand, the effective
diffusivity for the incompressible BC flow $\vel^I$ can be expressed as
(see, e.g., Ref.~\cite{BCVV})\,:
\be
\label{eddydiff}
\kappa^E_{\alpha\beta}=\kappa\delta_{\alpha\beta}
-{1\over 2}\left[ \langle v^I_{\alpha}s_{\beta}\rangle
+ \langle v^I_{\beta}s_{\alpha}\rangle \right],
\ee
where ${\bm s}$ has zero average value and satisfies
\be
\label{epsilon2}
\partial_t\,{\bm s}  + \left(\vel^I\cdot
{\bm \partial}\right) {\bm s} - \kappa\,
\partial^2 \,{\bm s} = -\vel^I .
\ee
The effective diffusivity, which is in general a second-order tensor,  
reduces for the BC flow to a scalar thanks to 
the following symmetries\,:
\be
\label{symm1}
x\mapsto y+{\pi\over 2}\,;\quad y\mapsto x-{\pi\over 2}\,;\quad
v^I_x\mapsto v^I_y\,;\quad v^I_y\mapsto v^I_x ,
\ee
and
\be
\label{symm2}
x\mapsto \pi-x\,;\quad y\mapsto y-\pi\,;\quad
v^I_x\mapsto -v^I_x\,;\quad v^I_y\mapsto v^I_y .
\ee
An immediate consequence of (\ref{symm1}) is that the diagonal 
components of the tensor in (\ref{eddydiff}) 
are equal and the symmetry (\ref{symm2}) implies 
that the non-diagonal components vanish. It follows that
\be
\label{kappay}
\kappa^E=\kappa^E_{xx}=\kappa^E_{yy}=- \langle v^I_{y}s_{y}\rangle ,
\ee
where $s_y$ is the zero-average solution of the equation
\be
\label{equaz}
\partial_t\,s_y  + \left( \vel^I\cdot
{\bm \partial}\right) s_y - \kappa\,
\partial^2 \,s_y = -\sin x .
\ee
%-----------------------------------------------------------------
\begin{figure}
\centerline{\psfig{figure=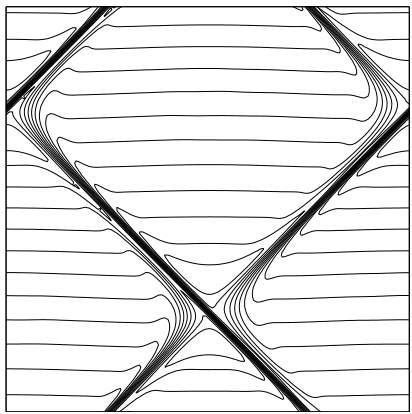,height=7cm}}
%,width=16cm}}
\vspace*{-0.7cm}
%\vspace{1cm}
\caption{Same as in Figure 2, with parameters
$\delta=0.05$ 
and $\kappa=0.001$.}
\vspace{0.3cm}
\label{kappa0.001}
\end{figure}
%-----------------------------------------------------------------

The crucial remark is now that equation (\ref{equaz}) coincides with 
the auxiliary equation (\ref{boh}) for the flow (\ref{perturbed})
at the dominant order in $\delta$.
When the solution of (\ref{boh}) is sought as
\be
\label{deltam}
m = m^{(0)} + \delta\, m^{(1)} + \delta^2\, m^{(2)} + \ldots ,
\ee
it is easy to check that $m^{(0)}=1$ and $m^{(1)}=-s_y$.
It follows from (\ref{Vb}), (\ref{kappasqrt}) and 
(\ref{kappay}) that the asymptotic 
behaviour of $\left( V_b\right)_y$ is
\be
\label{urra}
\left( V_b\right)_y \simeq {2\delta\over 0.749\sqrt{\pi}}\kappa^{1/2}
+ O(\delta^2).
\ee
Note that the behaviour in $\kappa$ of terms of higher order in $\delta$
is not under our control in this expansion.

The $x$-component of the 
ballistic velocity vanishes
for arbitrary $\delta$ and $\kappa$. 
Independently of the value of $\delta$, the
flow (\ref{perturbed}) has in fact the symmetry 
(\ref{symm2})\,:
\be
\label{symm3}
x\mapsto \pi - x\,;\qquad y\mapsto  y-\pi\,;\qquad v_x\mapsto -v_x\,;\qquad
v_y\mapsto v_y.
\ee
By using (\ref{symm3}) in (\ref{boh}), we immediately obtain
\be
\label{mmenom}
m(\pi-x\,;\,y-\pi)=m(x\,;\,y),
\ee
and
\be
\label{zero}
\left( V_b\right)_x=\langle v_x\,m\rangle =0.
\ee

The prediction (\ref{urra}) is compared in Table~1 to the values obtained 
by solving numerically the auxiliary equation (\ref{boh}) for
the flow (\ref{perturbed}). The equation has been solved by using 
pseudo-spectral methods \cite{GO} and the parameter $\delta=0.05$. The 
contour plots of the field $m$ are shown in Figs.~\ref{kappa0.01} and 
\ref{kappa0.001} for 
$\kappa=0.01$ and $\kappa=0.001$, respectively. The resolution is
$256\times 256$, which ensures that the boundary layers are properly
resolved.
\vspace{6mm}
\begin{center}
\begin{tabular}{|l|c|c|c|} \hline
\multicolumn{1}{|c}{\phantom{xxxx}} &
\multicolumn{1}{|c}{{\rm Spectral} $64^2$} &
\multicolumn{1}{|c|}{{\rm Asympt. Pred.}} \\
\hline
\multicolumn{1}{|c}{$\kappa$} &
\multicolumn{1}{|c|}{$\left( V_b\right)_y$} &
\multicolumn{1}{|c|}{$\left( V_b\right)_y$} \\
\hline \hline
$1.\times 10^{-1}$    &  $2.0\times 10^{-2}$ &  $2.4\times 10^{-2}$
 \\ 
$5.\times 10^{-2}$    &  $1.5\times 10^{-2}$ &  $1.7\times 10^{-2}$
 \\
$1.\times 10^{-2}$    &  $7.1\times 10^{-3}$ &  $7.5\times 10^{-3}$
 \\
$5.\times 10^{-3}$    &  $5.2\times 10^{-3}$ &  $5.3\times 10^{-3}$
 \\ \hline
\end{tabular}
\end{center}

\vspace{1mm}
\centerline{{\rm Table 1: The $y$-component of the 
ballistic velocity $\left( V_b\right)_y$ 
for the flow (\ref{perturbed}) with $\delta=0.05$.}}
\vspace{6mm}
It is evident that 
the quality of the prediction (\ref{urra}) 
improves when $\kappa$ is decreased.
This is 
in agreement with the asymptotic nature of the boundary-layer technique 
which has been used in the calculations.

\subsection{Random flows with short correlation times}
\label{s:Taylor}

It is known that the effective diffusivity for incompressible 
flow with short-correlation times can be calculated exactly (see, e.g.,
Ref.~\cite{BCVV}). 
The aim of this Section is to show that short correlation times
also allow to calculate the effective ballistic velocity for compressible 
flow.

Specifically, the velocity field is supposed to be a homogeneous,
Gaussian random process having zero average value and correlation function
\be
\label{correla}
\langle v_{\alpha}({\bm x},t)\,v_{\beta}({\bm 0},t')\rangle=
F_{\alpha\beta}({\bm x})\,\delta(t-t').
\ee

In order to calculate the average (\ref{Vb}) we can now use the formula
of Gaussian integration by 
parts (see, e.g., Refs.~\cite{UF,ZJ}). The average 
is then expressed as
\be
\label{deltaaa}
\Vb=\langle 
m\vel\rangle=\int\,d{\bm x}'\,dt'\langle\vel({\bm x},t)\vel({\bm x}',t')\rangle
\langle {\delta m({\bm x},t)\over \delta \vel ({\bm x}',t')}\rangle .
\ee
 From the auxiliary eq.~(\ref{boh}) and the $\delta$-correlation in time 
it immediately follows
\be
\label{Vbdelta}
\Vb = - \int_0^{\infty} \langle \vel ({\bm x},t)
\left({\bm \partial}\cdot{\bm v}\right)({\bm x},0)\rangle \,dt ,
\ee
where the stochastic differential equation (\ref{boh}) 
is interpreted \`a la Stratonovich since we are interested in the physical
limit of short correlation times.
Eq.~(\ref{Vbdelta}) had already been derived 
in Ref.~\cite{Israel} using path-integral
methods. Note that the average appearing in (\ref{Vbdelta}) vanishes for
incompressible flow. It also vanishes for parity-invariant and isotropic
flow, in agreement with the results of the next Section.

An alternative derivation of (\ref{Vbdelta}) 
is to 
neglect the diffusive term in
the auxiliary equation (\ref{boh}) and 
integrate on the characteristics\,:
\be
\label{esponenz}
m({\bm x},t) = {\exp \left(-\int_{-\infty}^t({\bm \partial}\cdot
\vel)({\bm x}({\bm a},s)\,;\,s)\,ds\right)\over
\langle \exp \left(-\int_{-\infty}^t({\bm \partial}\cdot \vel)(
{\bm x}({\bm a},s)\,;\,s)\,ds\right)\rangle } .
\ee
The characteristics are the Lagrangian trajectories defined by the 
equation
\be
\label{determ}
{d{\bm x}({\bm a},t)\over dt} = {\bm v}({\bm x}({\bm a},t)\,;\,t),
\ee
where ${\bm a}$ is the initial position of the particle. 
In (\ref{esponenz}) the integration has to made on the trajectory such that 
${\bm x}({\bm a},t)={\bm x}$. Inserting (\ref{esponenz}) into (\ref{Vb})
we obtain an expression of the ballistic velocity as time integral
of Lagrangian correlations. This is the equivalent of Taylor's formula
for effective diffusivities. The presence of the exponential makes
it however very difficult to use. 
When the flow has a short correlation time one can however simplify 
(\ref{esponenz}) by exploiting the fact that Lagrangian and Eulerian 
statistics tend to coincide. Using again Gaussian integration by parts 
(\ref{Vbdelta}) immediately follows.

\section{General conditions for the absence of ballistic $\,\,$
transport}
\label{s:absence}

The aim of this Section is to identify some 
general classes of flows where ballistic transport is absent and pure
diffusion takes place.

A first class where $\Vb=0$ is the one of incompressible flow. This has 
already been
shown by (\ref{incompressible}) in Section~\ref{s:ball}. 

Let us then consider parity-invariant flow, i.e.
velocity fields having at least one center of symmetry according to the
definition (\ref{parity}) in Section~\ref{s:formul}. 
It follows from (\ref{parity}) that the solution $m$
of (\ref{boh}) having unit average value satisfies
\be
\label{inverti}
m(-{\bm x},t) = m({\bm x},t).
\ee
In the random case, the statistics is invariant under 
the operations
\be
\label{parite}
{\bm x}\mapsto -{\bm x}\,;\qquad {\bm v}\mapsto -{\bm v}\,;\qquad
m\mapsto m.
\ee
An immediate consequence of (\ref{parity}) and (\ref{inverti}) (or
(\ref{parite}) for the random case) is the vanishing 
of the ballistic velocity $\Vb=0$.

\begin{figure}
\centerline{\psfig{figure=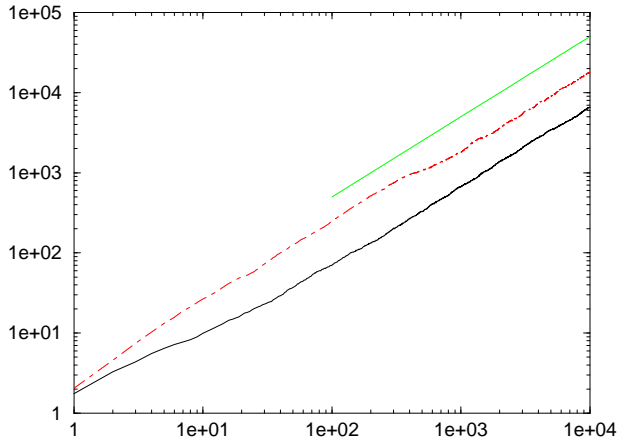,height=7cm}}
%,width=16cm}}
\vspace*{-0.7cm}
%\vspace{1cm}
\caption{The dispersions $\langle x^2(t)\rangle$ (lower curve) 
and $\langle y^2(t)\rangle$
for the flow (47). The straight line is proportional to $t$.}
\vspace{0.3cm}
\label{fig?}
\end{figure}

To give a direct confirmation of these considerations we present in
Fig.~\ref{fig?} the dispersions for the flow
\be
\label{piede}
\vel = (\sin x\,;\,\sin x + \sin 3x).
\ee
This flow depends on a single coordinate as discussed 
in Section~\ref{s:parallel}. Unlike 
(\ref{simple}), the flow (\ref{piede}) has however the center of symmetry 
$x=\pi$. It follows then that ballistic transport is forbidden. The
difference is best appreciated by comparing Figs.~1 and \ref{fig?}. 
In the first case  $\langle y^2(t)\rangle\propto t^2$ while 
in Fig.~\ref{fig?} $\langle y^2(t)\rangle\propto t$, 
corresponding to diffusive transport.

%-----------------------------------------------------------------
%-----------------------------------------------------------------

We consider now homogeneous and 
isotropic random flow. These flows are defined,
According to the definition in \cite{MY}, these flows are such that their
statistical properties are unaffected by translations and/or rotations
accompanied by a simultaneous rotation of the laboratory frame.
Note that, unlike the definition in Ref.~\cite{MY}, we distinguish
here between isotropy and statistical parity-invariance, e.g. 
a 3D helical flow can be isotropic even if it is not parity-invariant.
Since no preferential direction but $\vel$ appears in (\ref{boh}),
the statistical properties of the couple $(m,\vel)$ are invariant under
rotations. Homogeneity and the absence of a preferential direction imply then
that the single-point correlation
\be
\label{single}
\Vb = \langle \vel({\bm x},t)\,m({\bm x},t)\rangle =0.
\ee

In practice, information on isotropy or parity-invariance of the
flow may not be available. 
To check whether ballistic transport is present 
one could then use the
expansion in powers of ${1\over \kappa}$
derived in Appendix B. To leading order, this expansion shows that
\be
\label{marco1}
\Vb\ =\ -{1\over \kappa}\cdot
\langle {\bm v}{\cal H}^{-1}\left({\bm \partial}
\cdot\vel\right)\rangle +\ O\left({1 \over \kappa^2}\right)\ ,
\ee
where ${\cal H}^{-1}\ =\  \left(\partial_{\tau}-\partial^2 \right)^{-1}$ is
the Green's function of the heat diffusion operator. For time-independent
flows, the latter expression reduces to 
\be
\label{marco2}
\Vb\ =\ -{1\over \kappa}\cdot
\langle {\bm v}^I\,\phi
\rangle +\ O\left({1 \over \kappa^2}\right)\ ,
\ee
where ${\bm v}^I$ represents the incompressible component of the velocity
field and $\phi$ the potential of the irrotational component. 
The averages in the asymptotic expressions
(\ref{marco1}) or (\ref{marco2}) are easy to 
measure and 
can be used to estimate whether the ballistic
drift is present or not.

Finally, we investigate time-independent potential 
velocity fields, i.e. satisfying
\be
\label{rotzero}
\vel({\bm x}) = -{\bm \partial}\phi({\bm x}),
\ee
where $\phi$ is the potential.
The stationary solution of (\ref{boh}) for this class of flows 
is the Boltzmann distribution (see, e.g., Ref.~\cite{Risken})\,:
\be
\label{Gibbs}
m={\exp\left(-{\phi\over \kappa}\right)\over 
\langle \exp\left(-{\phi\over \kappa}\right)\rangle }.
\ee
It follows from periodicity (or homogeneity) that
\be
\label{merde}
\Vb \propto \,\langle\vel\exp\left(-{\phi\over \kappa}\right)\rangle
\,=\,\kappa\langle {\bm \partial}\left(\exp\left(-{\phi\over \kappa}\right)\right)
\rangle \,=\,0.
\ee
Ballistic effect is therefore impossible  
for time-independent potential flow. 
Note that the condition of time-independence is essential 
for this result. Indeed, if $\phi$ depends on time the 
solution of (\ref{boh}) is not
the Boltzmann distribution.
Furthermore, let us consider the one-dimensional flow
\be
\label{esempio}
v(x,t)=\cos (x-\kappa\,\omega\, t) + a\cos (x+\kappa\, \omega\, t).
\ee
Using (\ref{gener}) in Appendix~A, it is easy to check that 
the first coefficient of the perturbative expansion in $1/\kappa$
of the ballistic velocity is
\be
\label{timedep}
V_b=\left({\omega\over 1+\omega^2} {1-a^2\over 2}\right){1\over \kappa}
+O\left({1\over \kappa^2}\right). 
\ee
The fact that (\ref{timedep}) does not vanish illustrates the fact that
ballistic transport is possible for time-dependent potential flow.
One can also easily check that the expression (\ref{Vbdelta}), valid for
compressible $\delta$-correlated flow,
does not generally vanishes for potential flow.

\section{Effective diffusivities for compressible flow}
\label{s:diff}

In the previous Section we have considered advective effects
in large-scale transport by compressible flow. It is physically 
clear that diffusive effects will also be present. In order to 
systematically treat them, multi-scale techniques can again be used. 
To capture both advective and diffusive effects, two 
independent slow times 
$T_1=\epsilon t$ and $T_2=\epsilon^2 t$ are 
needed. The solvability condition
at order $\epsilon$ gives the advective effects, as shown in 
Section~\ref{s:ball},
while at order $\epsilon^2$ we also include diffusive effects (see 
Ref.~\cite{Complex}). For the flows where the ballistic velocity $\Vb$
vanishes, no advective time is needed. 
The formalism is then the same as in Section~\ref{s:ball}, 
with the only
difference that the slow time is now defined as $T=\epsilon^2 t$.
Since there are just technical differences between the two
cases $\Vb\neq 0$ and $\Vb=0$, we shall treat in detail the latter 
and give only the final results for
the former.

As in Section~\ref{s:ball} we denote the fast variables by
${\bm x}$, $t$ and the slow space variable is 
defined as ${\bm X}=\epsilon{\bm x}$.
The analogous of equation (\ref{slow}) for the derivatives
is now\,:
\be
\label{slow2}
\partial_i \mapsto \partial_i + \epsilon \nabla_i \,\,\,,
\qquad \partial_t \mapsto \partial_t + \epsilon^2 \partial_T \,\,\,,
\ee
The solution $\Theta$ of the Fokker-Planck equation (\ref{FP})
is again sought in the form of a series in $\epsilon$.
When (\ref{slow2}) and (\ref{expansion}) are inserted
into (\ref{FP}), a hierarchy of equations of the form (\ref{proto}) is
recovered. For the analysis of turbulent diffusion the following
three equations are needed\,:
\begin{eqnarray}
\label{ger1}
{\cal A} \Thzero &=& 0\, , \\
\label{ger2}
{\cal A} \Thuno &=& - \vel\cdot\nabla \Thzero
+2\kappa{\bm \partial}\cdot\nabla\Thzero\, , \\
\label{ger3}
{\cal A} \Theta^{(2)} &=& -\partial_T \Thzero - \vel\cdot\nabla \Thuno
+2\kappa{\bm \partial}\cdot\nabla\Thuno + \kappa\nabla^2\Thzero .
\end{eqnarray}
The operator ${\cal A}$ is defined by (\ref{proto}). Using  
the linearity 
of (\ref{ger1}), we can express the solution as in (\ref{measure}), 
where the field $m$
satisfies (\ref{boh}). We can then insert (\ref{measure}) into (\ref{ger2}).
Using again the linearity of the operator ${\cal A}$, we obtain:
\be
\label{evai}
\Thuno = \chiv\cdot\nabla\langle \Thzero\rangle + \, m\langle\Thuno\rangle .
\ee
where the vector field $\chiv$ satisfies
\be
\label{beep}
\partial_t \chiv + \underline{{\bm \partial}}\cdot
\left(\underline{\vel} \chiv \right) - \kappa\,
\partial^2 \chiv = -\vel\,m + 2 \kappa\, {\bm \partial}\,m .
\ee
The underline in the l.h.s. of (\ref{beep}) 
is meant to make it clear that in the scalar product the components of 
${\bm \partial}$ have to be contracted only with those of ${\vel}$, and not
of $\chiv$. 
As expected, the solvability condition 
for (\ref{beep}) requires $\Vb=0$ (which 
we have supposed to be true). In the cases when $\Vb$ does not vanish, 
the calculations with the two-times formalism give the following modified
equation 
\be
\label{beep2}
\partial_t \chiv +\underline{{\bm \partial}}
\cdot\left(\underline{\vel} \chiv \right) - \kappa\,
\partial^2 \chiv = -m\,\left(\vel-\Vb\right) + 2 \kappa\, {\bm \partial}\,m ,
\ee
where $\Vb$ is given by (\ref{Vb}). Eq.~(\ref{beep}) is clearly 
a particular case of (\ref{beep2}). 

The equation for the large scales dynamics
emerges as the solvability condition for (\ref{ger3})
\be
\label{diffusione}
\partial_T\langle\Thzero\rangle = 
\kappa^E_{\alpha\beta}\nabla_{\alpha}\nabla_{\beta}
\langle\Thzero\rangle ,
\ee
where the effective diffusivity tensor is
\be
\label{eddykappa}
\kappa^E_{\alpha\beta} = \kappa \delta_{\alpha\beta}
-{1\over 2}\left[\langle v_{\alpha}\chi_{\beta}\rangle + 
\langle v_{\beta}\chi_{\alpha}\rangle
\right] .
\ee

When $\Vb\neq 0$, the effective equation for the mean value of the 
field becomes
\be
\label{diffusione2}
\partial_T\langle \Thzero\rangle + \Vb\cdot\nabla\langle\Thzero\rangle = 
\kappa^E_{\alpha\beta}\nabla_{\alpha}\nabla_{\beta}
\langle\Thzero\rangle,
\ee
where the effective diffusivity has the expression  
(\ref{eddykappa}) and the $\chiv$ field is the solution of (\ref{beep2}).

The 
diffusivity tensor $\kappa^E_{\alpha\beta}$ 
might of course be anisotropic, reflecting the fact
that the flow could be non-invariant under arbitrary rotations.
The term in the square parentheses in (\ref{eddykappa}) does not have
a definite sign for generic compressible flow. The quantity 
$\kappa^E-\kappa$ is in fact known to be positive for 
incompressible flow and in the next Section we show
that it is always negative for static potential flow (see also the following
example). 

There are some special
instances where the 
effective diffusivity can be calculated analytically, 
e.g. the flows of the class (\ref{stratified}). We shall not dwell on
the details here 
since the solution of (\ref{boh}) is the Boltzmann distribution
(\ref{Boltz}) 
and the solution of (\ref{beep}) is essentially the same as the one which
will be discussed in 
Section~\ref{s:oned}.
The final result is that $\kappa^E_{xx}$ is given by (\ref{periodo}), the
non-diagonal components vanish and 
\be
\label{kappayy}
\kappa^E_{yy} = \kappa + {\langle e^{\phi/\kappa}\rangle \over \kappa}
\langle\left(\psi-{\langle\psi e^{\phi/\kappa}\rangle
\over \langle e^{\phi/\kappa}\rangle}\right)^2\left({e^{\phi/\kappa}\over 
<e^{\phi/\kappa}>}\right)\rangle ,
\ee
which is clearly larger than $\kappa$. The function $\psi$ is defined by
\be
\label{psi}
\psi(x) = \int_0^x w(z)\exp\left(-{\phi(z)\over \kappa}\right)\,dz,
\ee
and it is still a periodic function since the ballistic 
velocity is supposed to vanish.

For generic compressible flow, the
solutions of the auxiliary equations (\ref{boh}) and (\ref{beep}) 
are not generally known and
one should then resort to numerical methods, as in the incompressible case
\cite{BCVV,Majda}.

\section{Static potential flow}
\label{s:deplete}

One of the results derived in Section~\ref{s:absence} 
is that no ballistic transport is possible for
static potential flow. In the presence of scale separation, transport 
by these flows is therefore always  
diffusive. This is rather intuitive from a physical point of view. 
Particles are indeed concentrated into the minima of the 
potential where the velocity however vanishes. For the flow in 
Section~\ref{s:parallel} one immediately realizes, for example, that  
without the streaming in the $\hat{y}$-direction due to the 
solenoidal part no ballistic advection would be possible. 
As a matter of fact, we expect that even diffusion should 
be very sensitive to the presence of potential wells. Let us indeed consider
a stagnation point ${\bm x}_0$ ($\vel({\bm x}_0)=0$) and the quantity
\be
\label{indica}
{\cal T}=\vel({\bm x})\cdot({\bm x}-{\bm x}_0).
\ee
When ${\cal T}$ is expanded
in a Taylor series, the first non-zero term is generically quadratic\,:
\be
\label{indica2}
{\cal T}\simeq S_{ij}({\bm x}-{\bm x}_0)_i\,({\bm x}-{\bm x}_0)_j
= {1\over 2}\left(\partial_i v_j + \partial_j v_i\right)
({\bm x}-{\bm x}_0)_i\,({\bm x}-{\bm x}_0)_j.
\ee 
In order to have strong trapping the velocity field should be
such that the particles are driven back to ${\bm x}_0$, i.e. 
${\cal T}< 0$ everywhere in a neighbourhood of ${\bm x}_0$.
This is by definition what happens in the minima of the potential $\phi$
for the flows (\ref{rotzero}). On the contrary, this strong form of trapping
is impossible for incompressible flow. The tensor $S$ 
is in fact traceless
and not all its eigenvalues can be negative\footnote{Note that a {\em weak}
form of trapping is however possible \cite{RHK}.}. The conclusion of
these simple heuristic arguments is that a depletion of transport
is expected for static potential flow. This expectation is
confirmed by the systematic results of the next Section showing that
the effective diffusivity is indeed smaller than the molecular diffusivity.

\subsection{Depletion of transport}
\label{s:traps}

For static potential velocity fields
of the class (\ref{rotzero}) the calculation of the effective diffusivity
can be reduced to the solution of one auxiliary equation only.  
We can then show that diffusive transport is always depleted.

The solution of the first auxiliary equation (\ref{ger1})
is the Boltzmann distribution\,:
\be
\label{sol1}
\Thzero=\langle \Thzero\rangle 
({\bm X},T)\,\,{\exp\left(-{\phi\over \kappa}\right)\over 
\langle \exp\left(-{\phi\over \kappa}\right)\rangle }.
\ee
Inserting (\ref{sol1}) into (\ref{beep}), we obtain\,:
\be
\label{gamma}
\partial_t \Xiv + \underline{{\bm \partial}}
\cdot\left(\underline{\vel}\, \Xiv \right) - \kappa\,
\partial^2 \Xiv = \kappa\, {\bm \partial}\left(
\exp \left({-{\phi\over \kappa}}\right)\right),
\ee
where $\Xiv=\chiv\langle\exp\left(-{\phi\over \kappa}\right)\rangle$ 
and has zero
average value.
It is then convenient to define
\be
\label{chi}
\Upsv = \exp\left({\phi\over \kappa}\right) \Xiv  .
\ee
The equation for $\Upsv$ is easily derived from (\ref{gamma})
\be
\label{zeta}
\partial_t \Upsv - \left( \vel \cdot {\bm \partial}\right) 
\, \Upsv - \kappa\,
\partial^2 \Upsv = \vel .
\ee

We can now recast the averages appearing in the expression 
of the effective diffusivity  (\ref{eddykappa}) in a more convenient form.
Let us indeed multiply the components $\alpha$ and $\beta$ of (\ref{zeta}) 
by $\Xi_{\beta}$ 
and $\Xi_{\alpha}$, respectively. By summing, taking
the average and using the definitions of $\Xiv$ and 
$\Upsv$, we obtain after some algebra\,:
\be
\label{equality}
-{1\over 2}\left[\langle v_{\alpha}\,\chiv_{\beta}\rangle 
+\langle v_{\beta}\,\chiv_{\alpha}\rangle
\right] = - \kappa\langle \exp\left(-{\phi\over \kappa}\right)
({\bm \partial}\Upsilon_{\alpha})\cdot({\bm \partial}
\Upsilon_{\beta})\rangle\,/ \langle\exp\left(-{\phi\over \kappa}
\right)\rangle.
\ee
In order to derive this equality we have used the periodicity (or 
homogeneity) of $\vel$. 
It is evident from (\ref{equality}) 
that the correction to the molecular value which appears 
on the r.h.s. of (\ref{eddykappa}) is negative definite. This proves
that transport is always depleted. 

Let us then show that, despite of the fact that the correction due to the
flow ${\bm v}$ is negative, the whole effective diffusivity 
$\kappa^E_{\alpha\beta}$ is still positive, i.e.
the absolute 
value of the correction is always smaller than $\kappa$. Inserting 
the expression of $\chiv$ in terms of  
$\Upsv$ into (\ref{eddykappa}) and integrating by parts we have
\be
\label{deltak}
| \delta \kappa|\equiv \left(
\kappa \delta_{\alpha\beta}-\kappa^E_{\alpha\beta}\right)n_{\alpha}n_{\beta}
=\kappa {\langle\exp\left(-{\phi\over \kappa}\right)
\partial_{\parallel}(\Upsv\cdot\hat{n})\rangle\over
\langle \exp\left(-{\phi\over \kappa}\right)\rangle },
\ee
where $\hat{n}$ is a generic direction and $\partial_{\parallel}=
(\hat{n}\cdot{\bm \partial})$. 
Using the Schwartz inequality leads to 
\be
\label{chain}
|\delta\kappa | \le \kappa {\langle\exp\left(-{\phi\over \kappa}\right)
(\partial_{\parallel}(\Upsv\cdot\hat{n}))^2\rangle^{1/2}\over
\langle\exp\left(-{\phi\over \kappa}\right)\rangle^{1/2}}
\le
\kappa {\langle\exp\left(-{\phi\over \kappa}\right)
({\bm \partial}(\Upsv\cdot\hat{n}))^2\rangle^{1/2}\over
\langle\exp\left(-{\phi\over \kappa}\right)\rangle^{1/2}}
=\sqrt{\kappa |\delta\kappa |}.
\ee
Equation (\ref{equality}) has been used to derive the last equality.
The consequence of (\ref{chain}) is that the effective diffusivity tensor
$\kappa^E_{\alpha\beta}$ is indeed 
positive definite, in agreement with the 
known fact that the original Fokker-Planck equation is stable \cite{Risken}.

We have then shown that a static potential flow always leads
to depletion of transport. This should be contrasted to the case of 
incompressible flow, where transport properties are always enhanced. 
Remark that for {\it time-dependent}
potential flow the result just proved does not generally hold.
The time-dependence can in fact destroy the trapping effects due to
potential wells and lead to an
enhancement of transport. One can easily provide an explicit example 
by taking again the flow (\ref{esempio}) with $a=1$ 
(the flow is then parity-invariant and
no ballistic effect is present). The analytic expression of the effective
diffusivity is not known. For our aims it is however enough to
perform  
an expansion for large $\kappa$ quite similar 
to the one presented in Appendix~A. 
The first term of the expansion of $\kappa^E$ is obtained by some
simple algebra\,:
\be
\label{posneg}
\kappa^E=\kappa + {1\over \kappa}{\omega^2-3\over (\omega^2+1)^2}
+O(1/\kappa^2).
\ee
The fact that $\kappa^E$ can be either larger or smaller than $\kappa$
indicates that no general rule can be expected. Enhancement or
depletion of transport properties are both possible, depending on
the detailed structure of the flow.

\subsection{The one-dimensional case}
\label{s:oned}

The calculation of the effective diffusivity has been reduced to the 
solution of the auxiliary equation (\ref{gamma}). While in the 
multi-dimensional case no general solution is known, the equation
can be solved analytically in 1D. The reason for this 
simplification is that in one dimension
the order of the equation can be lowered,
thus reducing it to a first-order differential equation. We shall consider 
in detail the case of periodic, deterministic flow. The random case
is handled similarly. 

It follows from equation (\ref{zeta}) that 
\be
\label{primord}
{d\over dx}(\Upsilon +x)= c\exp\left( {\phi\over \kappa}\right) ,
\ee
where $c$ is a constant and $x$ is the spatial coordinate. 
One more integration leads to
\be
\label{onemore}
\Upsilon = c\,\int_0^x \exp\left({\phi\over \kappa}\right)
\,dy \,\,+\,\, c^{\prime} \,-\, 
x .
\ee
The constants $c$ and $c^{\prime}$ are fixed by imposing the conditions that
the function $\chi$ appearing in (\ref{chi}) 
is periodic and has zero average. For the calculation
of $\kappa^E$ only the value of $c$ turns out to be actually needed.
By imposing the periodicity of (\ref{onemore}), we finally obtain 
the expression of the effective diffusivity\,:
\be
\label{periodo}
\kappa^E = {\kappa\over \langle\exp({\phi\over \kappa})\rangle
\langle\exp(-{\phi\over \kappa})\rangle }.
\ee
It follows from (\ref{periodo})
that for small values of the molecular diffusivity the ratio
$\kappa^E/\kappa$ is exponentially small and 
has the structure of an Arrhenius factor. The term in 
(\ref{periodo}) is indeed
proportional to the probability for a particle  
to jump out of a well of depth
equal to the difference between the absolute maximum and minimum of the 
potential (see e.g. Section~5.10 in Ref.~\cite{Risken}). The mechanism of
transport is thus essentially a random 
walk between the minima of the potential.
Particles are trapped for very long times in the bottom of potential
wells. Because of large fluctuations in the noise, they can occasionally
jump out of a minimum falling into the adjacent one. The same
mechanism evidently  works also in the multidimensional case
and is responsible for the depletion of transport found in 
Section~\ref{s:traps}. 

For Gaussian random flow 
the averages in (\ref{periodo}) can be easily calculated\,:
\be
\label{Gaussian}
{\kappa^E\over \kappa}=\exp \left(-{\langle\phi^2\rangle
\over \kappa^2}\right).
\ee
Note that for transport to be diffusive at sufficiently large scales,
the condition
\be
\label{finito}
\langle\phi^2\rangle\quad < \infty ,
\ee
must be satisfied. When 
(\ref{finito}) is not satisfied, the hypothesis of scale separation
breaks down. The small-scale typical time for relaxation to local equilibrium 
is indeed of
the order of the typical Arrhenius time needed to jump out of the wells. 
When (\ref{finito}) is violated, this time is divergent because 
wells of arbitrary depth are present. 
There is therefore no time which is much larger than the typical times
of the small-scale dynamics.
Standard diffusion is never observed and transport is sub-diffusive. 
An explicit example is provided by the case when $\phi$ is a Brownian 
motion. Ya.~Sinai found in Refs.~\cite{Sinai1,Sinai2}
that in this case the dispersion does not vary linearly with time
but  
\be
\label{Yasha}
\langle (x(a,t)-x(a,0))^2\rangle \sim (\log t)^4.
\ee
Condition (\ref{finito}) is not satisfied also when
$\phi$ is a fractional Brownian motion.
This problem together with other examples of anomalous subdiffusive behavior
and applications to statistical mechanics systems
are discussed in detail in Ref.~\cite{BG}.

\section{Conclusions}
\label{s:conclusions}

Turbulent transport by compressible flow has been investigated. For generic 
compressible flow having non-vanishing solenoidal and potential components,
ballistic transport typically takes place. For isotropic and/or 
parity-invariant flow the ballistic velocity vanishes and transport
is diffusive. The properties of the effective diffusivities are strongly
dependent on the compressibility of the flow. For incompressible flow
it is known that diffusion is always enhanced in the presence of small-scale
velocity fields. On the contrary, we show that for static potential flow  
the effective diffusivities are reduced with respect to their molecular value.
Trapping effects due to potential wells are responsible for this depletion.
For time-dependent potential flow and generic compressible flow,
depletion or enhancement of transport are both possible. No general rule can
be drawn and the type of transport depends on the precise structure
of the flow. 
An essential role is in particular 
played by the geometry of channels and stagnation points of the velocity 
field.

\newpage

\section*{Appendices}

\appendix
\renewcommand{\theequation}{\thesection.\arabic{equation}}
\section{Expansion of the ballistic velocity for large $\;\;\;\;$
molecular diffusivities}
\setcounter{equation}{0}
In the limit of large molecular diffusivities $\kappa$, 
the ballistic velocity
can be expressed as a perturbative series
\be
\label{espansione}
\Vb = \sum_{n=1}^{\infty}{{\bm c}^{(n)}\over \kappa^n} .
\ee
The vector coefficients ${\bm c}^{(n)}$ are defined by 
\be
\label{nsima}
{\bm c}^{(n)} =\langle\vel\, m^{(n)}\rangle ,
\ee
where the functions $m^{(n)}$ are obtained recursively as
\be
\label{recursion}
m^{(n+1)} = - {\cal H}^{-1}\,{\bm \partial}\cdot\left(
\vel\,m^{(n)}\right)\qquad
n\ge 1,
\ee
and $m^{(0)}=1$. The operator ${\cal H}$ is the heat operator 
\be
\label{heat}
{\cal H} = \partial_{\tau}-\partial^2 ,
\ee
and $\tau = \kappa t$. The recursion relation 
(\ref{recursion}) is simply obtained by applying the inverse of the heat 
operator to (\ref{boh}) and expanding in powers of $1/\kappa$. Note that
for periodic, deterministic flow no problem in the 
inversion of ${\cal H}$ is encountered. The kernel of the operator is
in fact made up of constants which however cannot show up in the r.h.s.
of (\ref{recursion}) because of the divergence operator. In the random
case the convergence of the integrals involved in the 
coefficients ${\bm c}^{(n)}$ 
depends on the specific behaviour of the correlation functions of $\vel$.
The first coefficient in (\ref{espansione}) can be easily calculated
\be
\label{gener}
{\bm c}^{(1)}=-\langle {\bm v}{\cal H}^{-1}\left({\bm \partial}
\cdot\vel\right)\rangle .
\ee
The velocity field $\vel$ can be generally decomposed as the sum of 
a potential and a solenoidal part
\be
\label{Hodge}
\vel = -{\bm \partial}\phi + \vel^{I},
\ee
where $\vel^{I}$ is divergence-free.
For time-independent flow, (\ref{gener}) can be then reduced to the 
interesting form
\be
\label{primo}
{\bm c}^{(1)} = -\langle \phi\,\vel^{I}\rangle ,
\ee
which illustrates the importance of correlations between the potential 
and the solenoidal components of the velocity field 
for ballistic transport.

\section{Some useful formulae}
\setcounter{equation}{0}

In this Appendix we list the formulae which are needed 
for the calculation in Section~\ref{s:examples} of the ballistic 
velocity for the
flow (\ref{simple}). The formulae are taken from \cite{GR}.
\be
\label{AppB1}
\int_0^\pi \exp(-\beta\cos x)\,\cos x\,dx = i\pi\,J_1(i\beta)\,;
\ee
\be
\label{AppB2}
\int_0^\pi \exp(-\beta\cos x)\,dx = \pi\,J_0(i\beta)\,;
\ee
\be
\label{AppB3}
J_n(iz) = i^n I_n(z)\,;
\ee
\be
\label{AppB4}
I_n(z)=\sum_{k=0}^{\infty} {\left(z/2\right)^{2k+n}\over
k!\,(k+n)!}\,;
\ee
\be
\label{AppB5}
I_n(x) \sim {e^x\over \sqrt{2\pi x}}\sum_{k=0}^{\infty}
{(-1)^{k}\over (2x)^k}{\Gamma(n+k+1/2)\over k!\,\Gamma(n-k+1/2)}\,;
\ee
\be
\label{AppB6}
I_n(-x) \sim {e^{x}\over \sqrt{2\pi x}}e^{in\pi}\sum_{k=0}^{\infty}
{ (-1)^k\over (2x)^k}{\Gamma(n+k+1/2)\over k!\,\Gamma(n-k+1/2)}\,.
\ee
The functions $J_n$ and $I_n$ denote the Bessel functions of real
and imaginary arguments, respectively. In both 
(\ref{AppB4}) and (\ref{AppB5}) the variable $x$ is supposed to be real,
large and positive. 
The two couples of formulae (\ref{AppB1}), (\ref{AppB2}) 
and (\ref{AppB5}), (\ref{AppB6}) correspond to
3.915 and 8.451 in \cite{GR}. The (\ref{AppB3}) and (\ref{AppB4}) are 
8.406 and 8.445.


\newpage

\begin{thebibliography}{99}
\bibitem{FHP}
U.~Frisch, B.~Hasslacher \& Y.~Pomeau,
{\it Phys. Rev. Lett.}, {\bf 56}, 1505
%-1508
, (1986).
\bibitem{Complex}
U.~Frisch, D.~d'Humi\`eres, B.~Hasslacher, P.~Lallemand,
Y,~Pomeau \& J.P.~Rivet,
{\it Complex Systems}, {\bf 1}, 649
%-707
, (1987); also reproduced in
{\it Lattice Gas Methods For Partial Differential Equations}.
Ed. G.D, Doolen, Addison-Wesley, pp. 77.
%-135.
\bibitem{HKM}
H.K.~Moffatt,
%``Transport effects associated with turbulence,''
{\it Rep. Prog. Phys.}, {\bf 46}, 621, (1983).
\bibitem{BO}
C.M.~Bender \& S.A.~Orszag, {\it Advanced Mathematical Methods 
for Scientists and Engineers} (McGraw-Hill, 1978).
\bibitem{BLP}
A. Bensoussan, J.-L. Lions \& G. Papanicolaou, {\it Asymptotic
Analysis for Periodic Structures}, (North-Holland, Amsterdam, 1978).
\bibitem{Piretal}
D.~Mc Laughlin, G.C.~Papanicolaou \& O.~Pironneau, {\it SIAM Journal of Appl.
Math.}, {\bf 45}, 780, (1985).
\bibitem{YZ}
Ya.B.~Zeldovich,
%``Exact solution of the problem of diffusion in a periodic velocity field,
%and turbulent diffusion'',
{\it Sov. Phys. Dokl.}, {\bf 27}, 10, (1982).
\bibitem{Shraiman}
B.I.~Shraiman,
%``Diffusive transport in a Rayleigh-B\'enard convection cell,''
{\it Phys. Rev. A}, {\bf 36}, 261, (1987).
\bibitem{Rosetal}
M.N.~Rosenbluth, H.L.~Berk, I.~Doxas \& W.~Horton, {\it Phys. Fluids},
{\bf 30}, 2636, (1987).
\bibitem{BCVV}
L.~Biferale, A.~Crisanti, M.~Vergassola \& A.~Vulpiani, {\it Phys. Fluids},
{\bf 7}, 2725, (1995).
\bibitem{Majda}
A.~Majda \& R.~McLaughlin,
%``The effect of mean flows on enhanced diffusivity in transport by
%incompressible periodic velocity fields'',
{\it Stud. Appl. Math,}, {\bf 89}, 245, (1993).
\bibitem{FP}
A.~Fannjang \& G.C.~Papanicolaou, {\it SIAM J. Appl. Math.}, {\bf 54},
333, (1994).
\bibitem{AM}
M. Avellaneda \& A. Majda,
%``An integral representation and bounds on the
%effective diffusivity in passive advection
%and turbulent flows,''
{\it Commun. Math. Phys.}, {\bf 138}, 339, (1991).
\bibitem{AV}
M.~Avellaneda \& M.~Vergassola, {\it Phys. Rev. E}, {\bf 52}, 3249, (1995).
\bibitem{FSS}
U. Frisch, Z.S. She, and P.L. Sulem, {\it Physica}, D{\bf 28}, 382, (1987).
\bibitem{Siv}
G.I.~Sivashinsky, {\it Physica}, D{\bf 17}, 243, (1985).
\bibitem{GVF}
S.~Gama, M.~Vergassola \& U.~Frisch, 
{\it J. Fluid Mech.}, {\bf 260}, 95, (1994).
\bibitem{MV}
M.~Vergassola, {\it Europhys. Lett}, {\bf 24}, 41, (1993).
\bibitem{BMV}
R.~Benzi, A.~Manfroi \& M.~Vergassola, {\it Europhys. Lett}, 
{\bf 36}, 367, (1996).
\bibitem{Risken}
H.~Risken, {\it The Fokker-Planck Equation}, (Springer-Verlag, 1984).
\bibitem{HKM78}
H.K.~Moffatt, {\it Magnetic Field Generation in Electrically 
Conducting Fluids}, (Cambridge Univ. Press, 1978).
\bibitem{Israel}
T.~Elperin, N.~Kleeorin \& I.~Rogachevskii, {\it Phys. Rev E},
{\bf 52}, 2617, (1995).
\bibitem{AschM}
N.W.~Aschcroft \& N.D.~Mermin, {\it Solid State Physics}, (Holt-Saunders, 
1976).
\bibitem{GR}
I.S.~Gradshteyn \& I.M.~Ryzhik, {\it Tables of Integrals, Series and
Products}, (Academic Press, 1965).
\bibitem{Dombre}
T.~Dombre, U.~Frisch, J.M.~Greene, M.~H\'enon, A.~Mehr \& A.M.~Soward,
%``Chaotic streamlines in the ABC flows,''
{\it J. Fluid Mech.}, {\bf 167}, 353 (1986).
\bibitem{GO}
D.~Gottlieb \& S.A.~Orszag. {\it Numerical Analysis of Spectral
Methods}, SIAM, (1977).
\bibitem{UF}
U.~Frisch, {\it Turbulence}, Cambridge Univ. Press, Cambridge, (1995).
\bibitem{ZJ}
J.~Zinn-Justin, {\it Quantum Field Theory and Critical Phenomena},
Clarendon Press, Oxford, Second Edition (1993).
\bibitem{MY}
A.S.~Monin \& A.M.~Yaglom, {\it Statistical Fluid Mechanics}, edited by
J.~Lumley, (MIT Press, Cambridge, Mass.) 1975.
\bibitem{RHK}
R.H.~Kraichnan, {\it Phys. Fluids}, {\bf 13}, 22, (1970).
\bibitem{Sinai1}
Ya.G.~Sinai, in {\it Lecture Notes in Physics}, Vol.~153, eds. R.~Schrader,
R.~Seiler and D.~Uhlenbrock, Springer, Berlin, (1981).
\bibitem{Sinai2}
Ya.G.~Sinai, {\it Theor. Prob. Appl.}, {\bf 27}, 247, (1982).
\bibitem{BG}
J.P.~Bouchaud \& A.~Georges, {\it Phys. Rep.}, {\bf 195}, 127, (1990).
\end{thebibliography}
\end{document}